\documentclass[conference]{IEEEtran}
\IEEEoverridecommandlockouts
\usepackage{cite}
\usepackage{setspace}
\usepackage[numbers]{natbib}
\usepackage{hyperref}
\usepackage{balance}
\usepackage{amsmath,amssymb,amsfonts}
\usepackage{algorithmic}
\usepackage{graphicx}
\usepackage{subcaption}
\usepackage{wrapfig}
\usepackage{textcomp}
\usepackage{xcolor}
\def\BibTeX{{\rm B\kern-.05em{\sc i\kern-.025em b}\kern-.08em
    T\kern-.1667em\lower.7ex\hbox{E}\kern-.125emX}}
\usepackage{tikz}
\newcommand\submittedtext{%
  \footnotesize This work has been submitted to the IEEE for possible publication. Copyright may be transferred without notice, after which this version may no longer be accessible.}

\newcommand\submittednotice{%
\begin{tikzpicture}[remember picture,overlay]
\node[anchor=south,yshift=10pt] at (current page.south) {\fbox{\parbox{\dimexpr0.65\textwidth-\fboxsep-\fboxrule\relax}{\submittedtext}}};
\end{tikzpicture}%
}
\begin{document}

\newcommand{\emailone}{\textsuperscript{1}}
\newcommand{\emailtwo}{\textsuperscript{2}}
\newcommand{\emailthree}{\textsuperscript{3}}
\newcommand{\emailfour}{\textsuperscript{4}}
\newcommand{\emailfive}{\textsuperscript{5}}
\newcommand{\emailsix}{\textsuperscript{6}}
\newcommand{\CS}{\textsuperscript{*}}
\newcommand{\IT}{\textsuperscript{\#}}

\nocite{*}
\title{Autoencoded Image Compression for Secure and Fast Transmission}

\author{
\IEEEauthorblockN{
Aryan Kashyap Naveen\emailone\IT, 
Sunil Thunga\emailtwo\CS, 
Anuhya Murki\emailthree\CS, 
Mahati Kalale\emailfour\CS, 
Shriya Anil\emailfive\CS
}
\IEEEauthorblockA{
\CS\textit{Dept. of Computer Science and Engineering}, 
\IT\textit{Dept. of Information Technology} \\
\textit{National Institute of Technology Karnataka, Surathkal, India} \\
\texttt{\emailone aryankashyap.221ai012@nitk.edu.in, 
\emailtwo sunilthunga.221cs252@nitk.edu.in,} \\
\texttt{\emailthree anuhyamurki.211cs213@nitk.edu.in, 
\emailfour mahatikalale.211cs138@nitk.edu.in,} \\
\texttt{\emailfive shriya.211cs259@nitk.edu.in}
}
}


\maketitle
\submittednotice
\begin{abstract}
With exponential growth in the use of digital image data, the need for efficient transmission methods has become imperative. Traditional image compression techniques often sacrifice image fidelity for reduced file sizes, challenging maintaining quality and efficiency. They also compromise security, leaving images vulnerable to threats such as man-in-the-middle attacks. This paper proposes an autoencoder architecture for image compression to not only help in dimensionality reduction but also inherently encrypt the images. The paper also introduces a composite loss function that combines reconstruction loss and residual loss for improved performance. The autoencoder architecture is designed to achieve optimal dimensionality reduction and regeneration accuracy while safeguarding the compressed data during transmission or storage. Images regenerated by the autoencoder are evaluated against three key metrics: reconstruction quality, compression ratio, and one-way delay during image transfer. The experiments reveal that the proposed architecture achieves an SSIM of 97.5\% over the regenerated images and an average latency reduction of 87.5\%, indicating its effectiveness as a secure and efficient solution for compressed image transfer.
\end{abstract}


\section{Introduction}
In the era of high-resolution imagery, efficient image compression techniques are crucial for reducing storage requirements and transmission times. Convolutional Autoencoders (CAEs) have emerged as a powerful tool for unsupervised learning and data compression.

This paper introduces a novel approach to image compression using CAEs centred around a unique composite loss function. Our method combines reconstruction loss with a residual component, addressing a key challenge in autoencoder-based compression: preserving fine details while achieving high compression ratios. This innovative loss function draws inspiration from recent advancements in image super-resolution, enabling our model to capture high-frequency information often lost in conventional compression techniques. The core components of our convolutional autoencoder have been detailed below.

\noindent\textbf{Encoder}: A convolutional neural network that progressively extracts features from the input image through convolution and pooling layers, encoding them into a compact latent space representation.

\noindent\textbf{Decoder}: A mirrored CNN architecture that upsamples the latent representation through deconvolutional layers, aiming to reconstruct the original image with high quality.

\begin{figure}[h]
    \centering
    \includegraphics[width=0.48\textwidth]{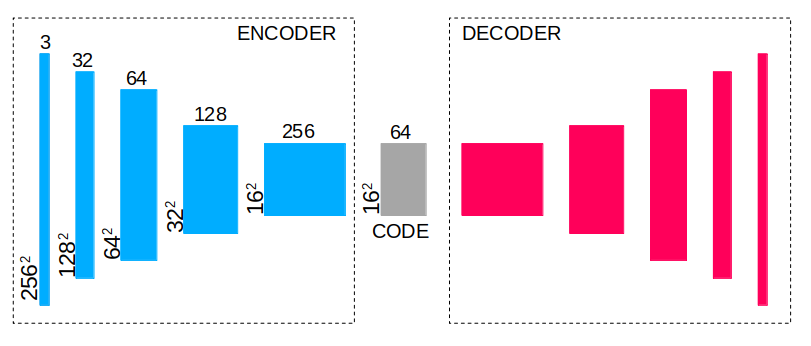}
    \caption{Autoencoder Architecture}
    \label{fig:autoencoder}
\end{figure}

The latent space representation generated by the encoder holds the key to inherent encryption. Unlike traditional compression techniques that operate directly on the image data, an autoencoder \cite{b11} compresses the image into a latent space representation that is not directly interpretable. The compressed representation is a product of the complex non-linear transformations learnt by the encoder's convolutional layers. Decoding this representation requires the specific architecture and weights of the trained decoder, making it inherently encrypted for anyone without access to the decoder network.

This paper presents an approach to secure image transmission \cite{b9} utilising a convolutional autoencoder trained on a designated image dataset, allowing it to learn the optimal representation of the images. Following training, the components of the autoencoder are separated. The encoder resides at the sender's host, while the decoder resides at the receiver's host. Images are compressed using the sender-side encoder, generating a latent space representation significantly smaller than the original image. The compressed data is then transmitted across the network using a client-server architecture. Using the above architecture, two key advantages are noticeable: The transmission times are significantly less than traditional methods, and the image is seemingly encrypted. The latent space representation itself acts as inherent encryption. Without the decoder present on the receiver's side, the compressed data is unintelligible, effectively preventing unauthorised access or manipulation during transmission. Hence, the distributed architecture offers a secure and efficient solution for image transmission.

The paper is structured as follows: Section II provides a literature review on the application of autoencoders for image compression as proposed in existing research. Section III outlines the datasets used for training and evaluation, while section IV delves into the architecture of the autoencoder proposed in this study. Section V elaborates on the training process of the autoencoder and introduces the novel loss function proposed. Section VI presents the results obtained and compares them with those of JPEG compression.  Lastly, Section VII offers concluding remarks and discusses potential future directions.

\section{Literature review}
Autoencoders have emerged as a promising tool for various tasks, including dimensionality reduction \cite{b15} and image compression \cite{b10}. Leveraging deep learning techniques, researchers have explored innovative approaches to enhance image retrieval and compression algorithms. 

Theis et. al \cite{b1} proposed a compressive autoencoder architecture for image compression, combining convolutional neural networks (CNNs) with efficient quantisation and entropy encoding methods. The method outperformed JPEG and JPEG 2000 in terms of PSNR (Peak Signal to Noise Ratio) and SSIM (Structural Similarity Index Metric), particularly at high bit rates. Subjective quality assessed via a mean opinion score (MOS) test showed Convolutional Autoencoders achieving higher scores than JPEG and JPEG 2000, with significant improvements at certain bit rates. The average bit rates of CAE compressed images were 0.24479, 0.36446, and 0.48596, respectively.

Cheng et. al \cite{b2} proposed a deep convolutional autoencoder-based image compression method that significantly reduced the bit rate compared to JPEG, demonstrating superior compression. Their approach integrated the Principal Component Analysis (PCA) to rotate feature maps, ensuring an energy-compact representation that enhanced entropy coding. This method handled image sizes up to $256\times256$, leading to notable improvements in the quality of reconstructed images. Their method outperforms conventional traditional image coding algorithms and achieves a 13.7\% BD-rate decrement compared to JPEG2000 on the Kodak database images.

Petscharnig et. al \cite{b3} explored the use of autoencoders for dimensionality reduction and feature fusion in image retrieval tasks, particularly in scenarios where training data is scarce or sensitive. The performance of autoencoder-based feature fusion was compared with single hand-crafted features using mean average precision (MAP) and precision at 10 (p@10) on UCID and SIMPLIcity datasets. The paper explored how different distance metrics (Euclidean, Manhattan, Jensen-Shannon divergence) affect retrieval performance and suggested a trade-off between dimensionality and information loss, proposing that low-dimensional representations can be used for fast approximate search, with re-ranking based on higher-dimensional features.

Mei et. al \cite{b22} proposed a scalable image compression method leveraging latent feature reuse and prediction to enhance compression efficiency. The approach employs an end-to-end auto-encoder framework, with a base layer providing the lowest resolution and multiple enhancement layers utilising previously learned latent features for improved quality and resolution. Results demonstrate that the method outperforms traditional scalable compression standards (SVC, SHVC) and single-layer methods in terms of both bit-rate efficiency and image quality. The approach's prediction mechanism further reduces redundancy, achieving better performance in scalable scenarios compared to existing methods. Evaluation of standard datasets reveals significant improvements in rate-distortion performance and efficiency. 

By combining these insights and advancements, the overall concept of using autoencoders for image compression and secure image transfer can be significantly enhanced.

\section{Dataset}
\subsection{Training Dataset}
The Stanford Dogs \cite{b4} and Animals 10 \cite{b5} datasets were used for training and evaluating the proposed convolutional autoencoder model. The Stanford Dogs dataset comprises 20,580 colour images spanning 120 different dog breeds. These images were initially gathered from the ImageNet\cite{b6} dataset. The Animals 10 dataset comprises 26,200 images belonging to 10 categories. These were obtained from Google Images.
All images were resized to 256×256 pixels in RGB format using bicubic interpolation to ensure uniform input size for the model. The images were then encoded as tensors with the shape (256, 256, 3). 
\subsection{Evaluation Dataset}
For evaluation, we used the Natural Images dataset \cite{b7}, which consists of images from various classes. Eight images were selected, one from each class, for testing.

\section{Architecture}
The proposed architecture consists of a convolutional autoencoder designed for progressive image compression. The encoder encodes the input image into a latent space representation, while the decoder aims to reconstruct the original image from this compressed form. The key novelty of our approach lies in the loss calculation, which combines residual loss and reconstruction loss. The residual loss, inspired by image super-resolution models\cite{b20} \cite{b21}, captures high-frequency details. Minimising the residual loss enables the model to quickly learn efficient representations. Meanwhile, the reconstruction loss encourages the autoencoder to develop a compact representation of the input image, facilitating effective compression. By jointly optimising these losses, our model achieves an optimal trade-off between image quality and compression ratio.
Existing methods typically rely on entropy coding and quantisation for compression. However, our approach achieves comparable compression results using solely an autoencoder architecture. The autoencoder learns to compress and reconstruct images in an end-to-end fashion, eliminating the need for traditional compression techniques.

\begin{figure}[h]
    \centering
    \includegraphics[width=0.48\textwidth]{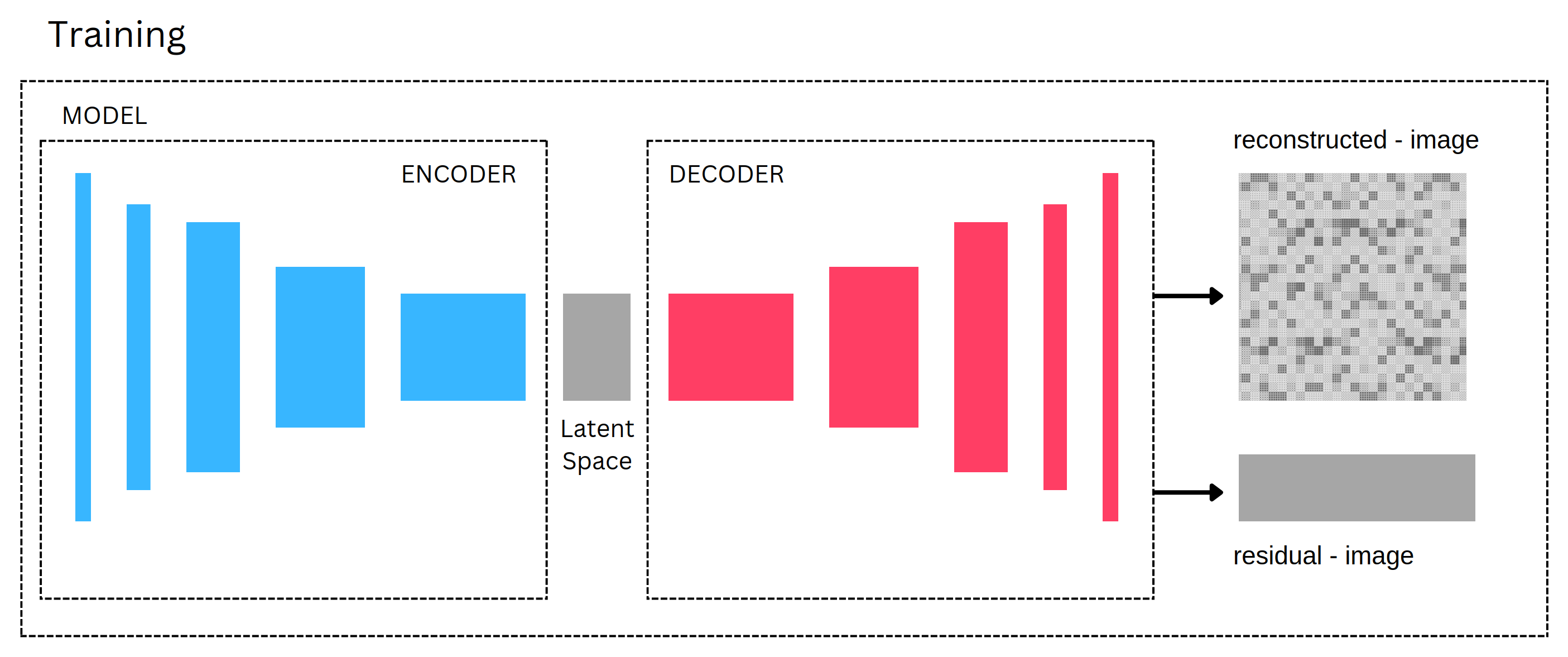}
    \caption{Model setup during training}
    \label{fig:training}
\end{figure}

\begin{figure}[h]
    \centering
    \includegraphics[width=0.48\textwidth]{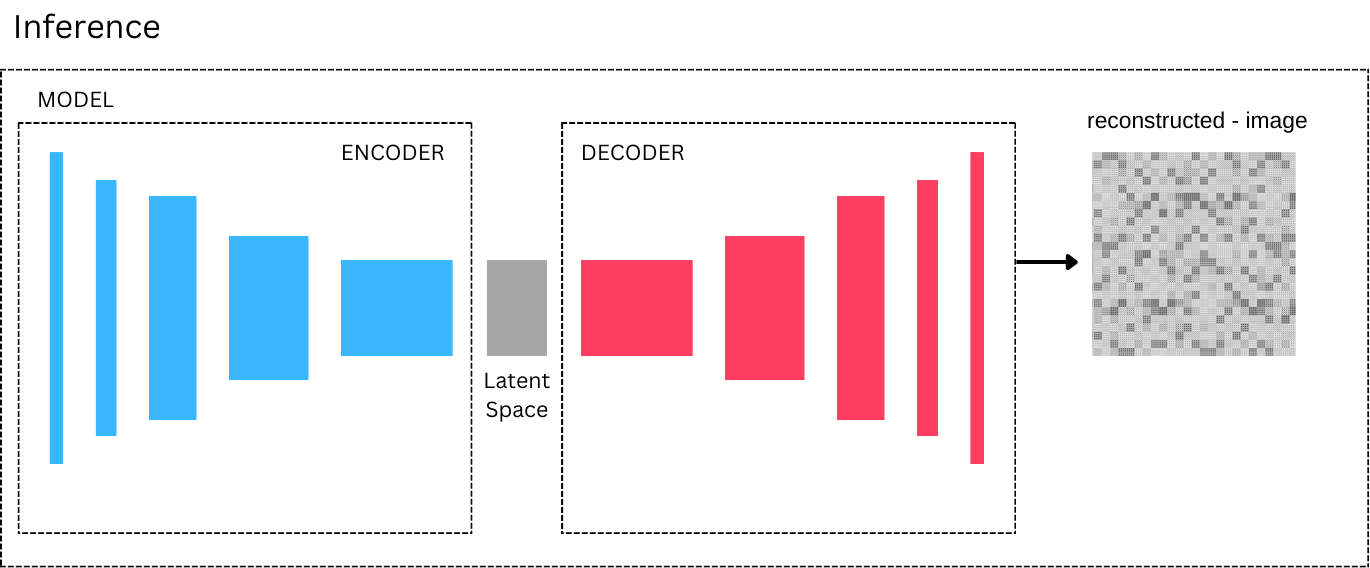}
    \caption{Model setup during testing}
    \label{fig:testing}
\end{figure}

\subsection{Encoder}
The encoder comprises four convolutional layers with ReLU activations, each succeeded by a $2\times2$ max-pooling operation for downsampling. The number of filters doubles after each convolutional layer, beginning with 32 in the first layer and reaching 256 in the fourth layer. This progressive increase in filters enables the encoder to capture increasingly complex and abstract features with greater depth. Lastly, a bottleneck convolutional layer having 64 filters is employed to transform the encoded representation to a lower-dimensional space.

\subsection{Decoder}
The decoder architecture is symmetrical to the encoder. It has four transpose convolutional (deconvolution) layers with ReLU activations to progressively upsample and reconstruct the original image. The number of filters in the decoder layers decreases by a factor of two after each layer, starting from 256 and ending with 32 filters. The final layer is a transposed convolutional layer with 3 filters and a sigmoid activation to produce the reconstructed image in the RGB colour space.

\begin{figure}[h]
    \centering
    \includegraphics[width=0.42\textwidth]{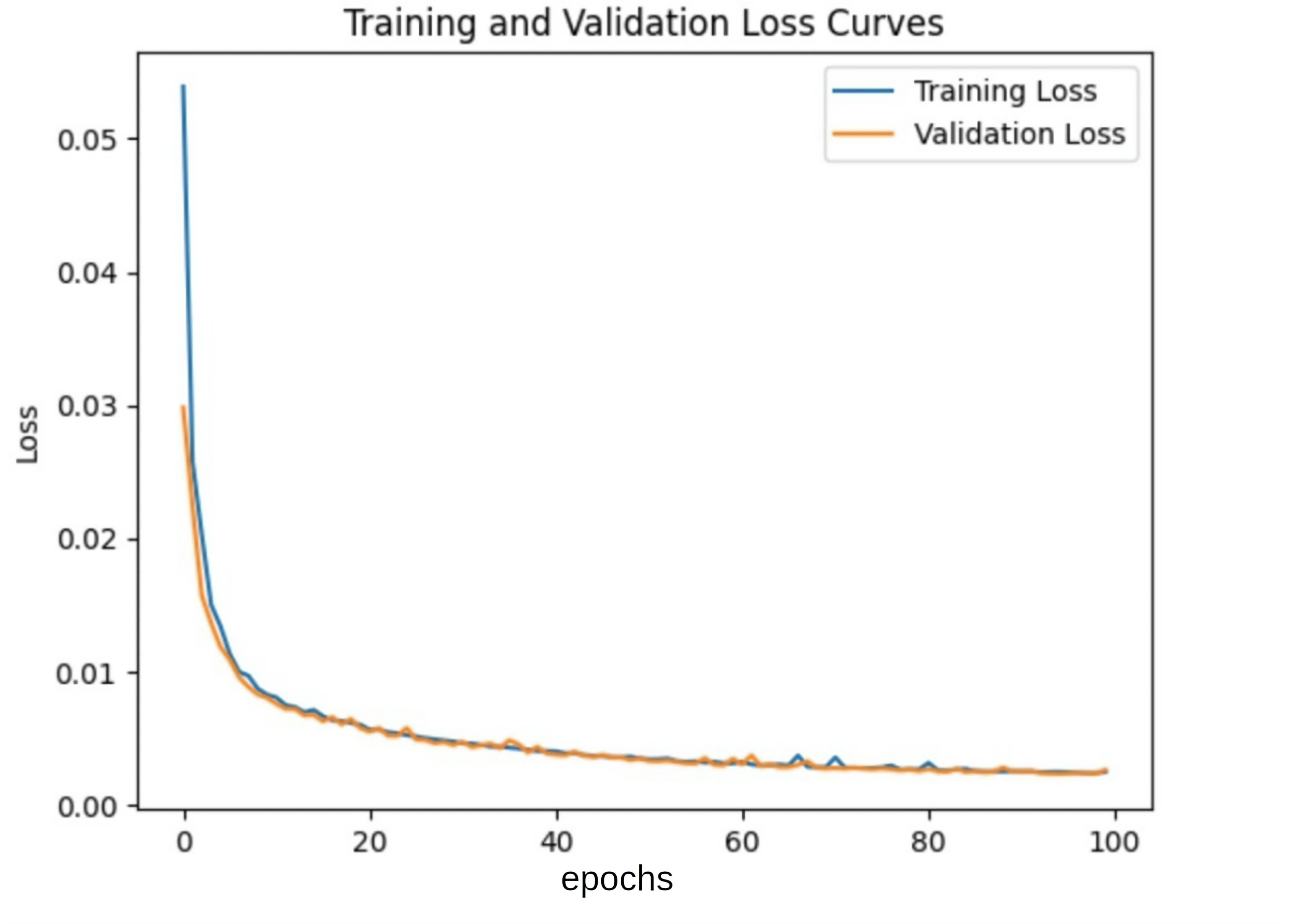}
    \caption{Loss vs Epochs during training}
    \label{fig:loss_vs_epochs}
\end{figure}

The progressive compression aspect of the model is achieved by allowing 
the decoder to output not only the reconstructed image but also a 
residual image during training, as shown in Figure \ref{fig:loss_vs_epochs}. 
This residual image represents the difference between the original input 
image and the reconstructed image, capturing the information lost during 
the compression and reconstruction process. Throughout training, the 
model minimises the information carried by the residual image. During 
inference, the model wrapper responsible for outputting the residual 
image is removed.

\section{Training}

\subsection{Model and Setup}
The training process employed a progressive compression method, integrating reconstruction and residual components to optimise the convolutional autoencoder for image compression. The encoder and decoder components were encapsulated within a model wrapper, allowing the model to output both the reconstructed and residual images during training. The residual image, representing the difference between the input and reconstructed images, facilitated higher reconstruction quality and reduced training times.

Training was conducted on two NVIDIA T4 GPUs, leveraging PyTorch's DataParallel module for parallel computation across both GPUs. This parallelisation enabled efficient training with larger batch sizes.

\vspace{-5pt}

\begin{align*}
\text{Forward Pass:} \quad
e &= \text{Encoder}(x) \\
d &= \text{Decoder}(e) \\
r &= x - d \\
(d, r) &= \text{Model}(x)
\end{align*}

\subsection{Hyperparameters}
Our experiment revealed that a batch size of 8 yielded the best results, as detailed in the results section. We used an \textbf{80/20} train/validation split. The Adam optimiser with an initial learning rate of 0.001 was utilised. A `reduce learning rate on plateau' scheduler was used to ensure convergence and prevent overfitting. This scheduler reduced the learning rate by a factor of 0.1 when the validation loss remained stagnant for ten successive epochs. Training was conducted for a total of 100 epochs. Additionally, the Early stopping mechanism halted training if the validation loss failed to improve over 20 epochs, thereby preventing overfitting.

\subsection{Loss Calculation}

Traditional autoencoders typically focus solely on reconstruction loss, which can lead to blurry outputs or loss of fine details. Our approach draws inspiration from image super-resolution and residual learning. 
The key insight was to explicitly model and minimise the residual information - the difference between the input and reconstructed image. By incorporating this residual loss alongside the traditional reconstruction loss, we encourage the model to learn an efficient compression strategy that preserves high-frequency details often lost in conventional methods.

The objective loss function guiding the training process was the sum of the reconstruction loss and residual loss, each calculated by the mean squared error (MSE). The reconstruction loss measured the difference between the input image and the image reconstructed by the decoder. The residual loss quantified the difference between the residual image and a target tensor of zeros. By minimising both the reconstruction error and the residual image, the final loss used for training enabled the model to learn an efficient compression strategy.

\
\begin{align*}
\text{Training Loop:} \quad
\mathcal{L}_r &= \text{Criterion}(d, x) \\
\mathcal{L}_i &= \text{Criterion}(r, \mathbf{0}) \\
\mathcal{L} &= \mathcal{L}_r + \mathcal{L}_i
\end{align*}
\vspace{-10pt}
\begin{align*}
\text{Key:}\\
x&: \text{Input image} \\
e&: \text{Encoded representation of the input image} \\
\text{Encoder}(\cdot) &: \text{Encoder function} \\
d&: \text{Decoded (reconstructed) image} \\
\text{Decoder}(\cdot) &: \text{Decoder function} \\
r&: \text{Residual image} \\
\text{Model}(\cdot) &: \text{Autoencoder model} \\
\mathcal{L}_r&: \text{Reconstruction loss} \\
\text{Criterion}(\cdot, \cdot) &: \text{Loss function (mean squared error)} \\
\mathbf{0}&: \text{Tensor of zeros, shape = residual tensor} \\
\mathcal{L}_i&: \text{Residual loss} \\
\mathcal{L}&: \text{Total loss } (\mathcal{L}_r + \mathcal{L}_i)
\end{align*}

\vspace{5pt}

\section{Results}
\subsection{Effect of the Residual Component}
The architecture was evaluated by comparing the performance of two models: one incorporating residual loss (progressive compression) and the other without it.

Training used a batch size of 256 and Distributed Data Parallel (DDP) for 20 epochs. After training, the models were tested on 8 natural images from different classes. The model with residual loss achieved a higher Peak Signal to Noise Ratio (PSNR) and Structural Similarity Index (SSIM) \cite{b8}, and lower Mean Squared Error (MSE), indicating better reconstruction quality.

\begin{table}[htbp]
\caption{Comparison between models with and without progressive compression}
\begin{center}
    \renewcommand{\arraystretch}{1.5} 
    \setlength{\tabcolsep}{10pt} 
    \begin{tabular}{|c|c|c|c|}
        \hline
        \textbf{Model} & \textbf{PSNR} & \textbf{SSIM} & \textbf{MSE} \\
        \hline
        Without PgIC & 29.5916 & 0.8869 & 0.0012 \\
        \hline
        With PgIC (proposed) & 31.3478 & 0.8969 & 0.0008 \\
        \hline
    \end{tabular}
\end{center}
\end{table}

\begin{figure}[h]
    \centering
    \begin{subfigure}{0.48\textwidth}
        \centering
        \includegraphics[width=\linewidth]{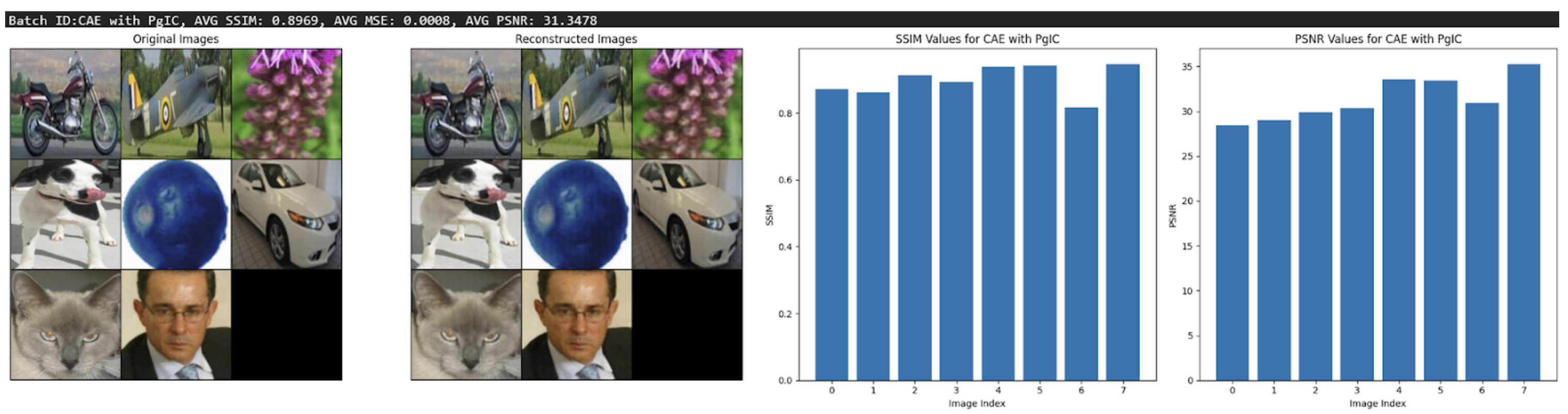}
        \caption{Results with Progressive Image Compression}
        \label{fig:with_residual_fft}
    \end{subfigure}%
    \hfill
    \begin{subfigure}{0.48\textwidth}
        \centering
        \includegraphics[width=\linewidth]{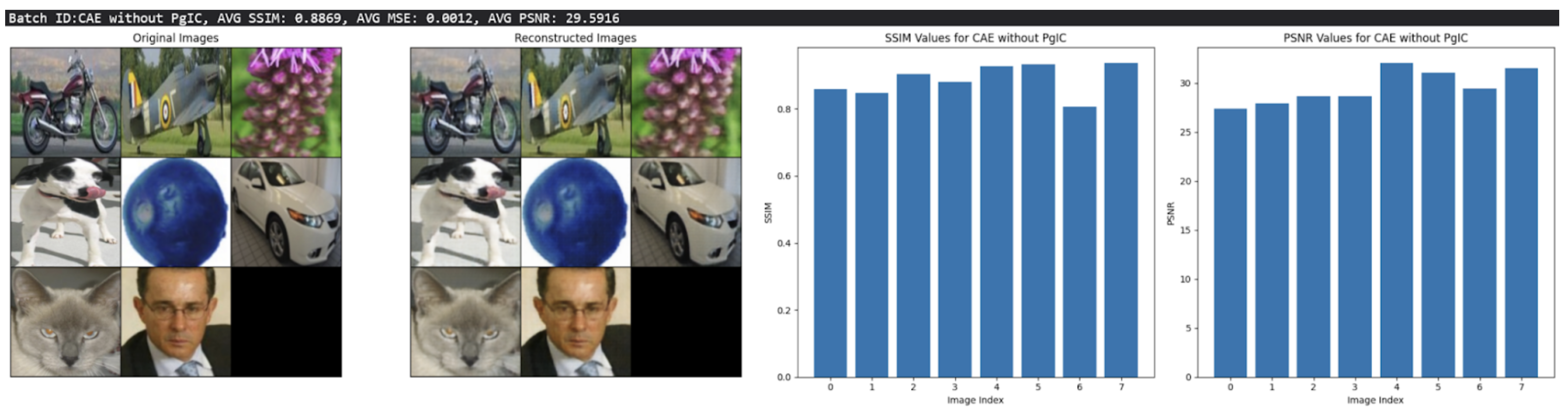}
        \caption{Results without Progressive Image Compression}
        \label{fig:without_residual_fft}
    \end{subfigure}
    \caption{Comparison of results with and without Progressive Image Compression}
    \label{fig:example}
\end{figure}

To further understand the impact of the residual loss component on the reconstruction quality, we conducted a frequency domain analysis. This analysis provides insights into how well each model preserves different levels of image detail, from broad structures to fine textures. We applied the 2D Fourier Transform to both the original and reconstructed images, allowing us to visualise their frequency components. The images used were passed to the model after training it for 30 epochs. We analysed three key visualisations, Original Frequency Spectrum, Reconstructed Frequency Spectrum and Frequency Difference Spectrum.
The frequency difference spectrum, in particular, highlights discrepancies between the original and reconstructed images across different frequency bands.

\begin{figure}[h]
    \centering
    \includegraphics[width=0.45\textwidth]{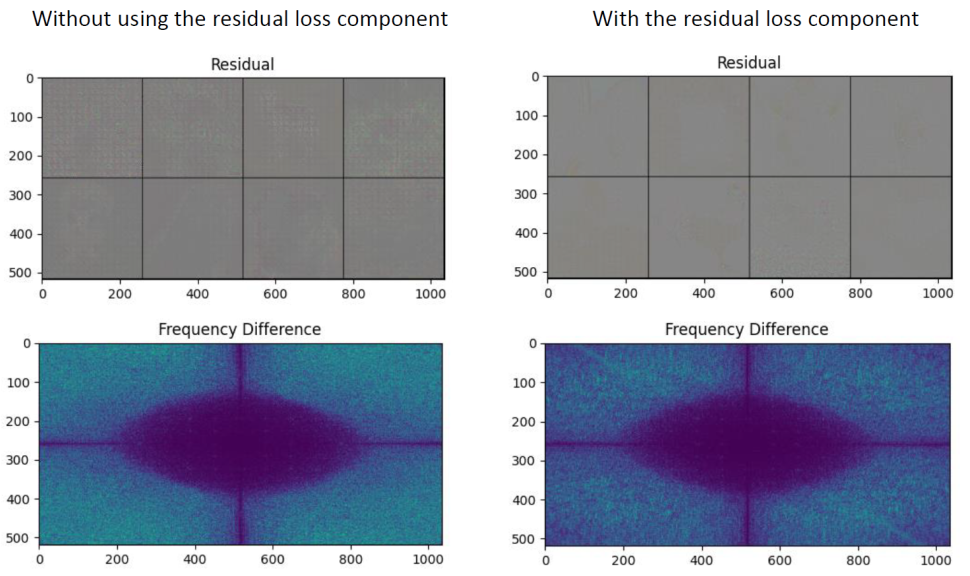}
    \caption{Frequency Domain and Residual Component Analysis}
    \label{fig:freq_analysis}
\end{figure}

Figure \ref{fig:freq_analysis} shows the frequency difference spectrum and output residual image for both models. The following observations can be made:

\begin{enumerate}
    \item \textbf{Low-Frequency Components:} Both models show strong preservation of low-frequency components, indicated by the dark centre in the difference spectrum. This suggests that both approaches maintain the overall structure of the images well.
    \item \textbf{Mid-Frequency Components:} The model with residual loss demonstrates better preservation of mid-frequency details. This is evident from the darker regions in the mid-range of its difference spectrum compared to the model without composite loss.
    \item \textbf{High-Frequency Components:} The most notable difference is observed in the high-frequency range. The model with residual loss shows a significantly darker outer region in the difference spectrum, indicating superior preservation of fine details and textures.
    \item \textbf{Residual Image:} In the residual images for both models, it is clear that the model which uses the composite loss function to retain more detail in the residual image.
\end{enumerate}

\subsection{Performance Evaluation Across Varied Batch Sizes using Progressive Image Compression (PgIC)}

\begin{table}[h]
\caption{Performance evaluation for different batch sizes}
\begin{center}
    \renewcommand{\arraystretch}{1.5} 
    \setlength{\tabcolsep}{15pt} 
    \begin{tabular}{|c|c|c|c|}
        \hline
        \textbf{Batch Size} & \textbf{PSNR} & \textbf{SSIM} & \textbf{MSE} \\
        \hline
        8 & 37.0432 & 0.9757 & 0.0002 \\
        \hline
        16 & 36.9585 & 0.9748 & 0.0002 \\
        \hline
        32 & 37.0554 & 0.9754 & 0.0002 \\
        \hline
        64 & 36.2745 & 0.9708 & 0.0003 \\
        \hline
        256 & 32.4127 & 0.9295 & 0.0006 \\
        \hline
    \end{tabular}
\end{center}
\end{table}

Selecting the optimal batch size is pivotal in training deep-learning models. Thus, we conducted experiments across a range of batch sizes, from 8 to 256. The evaluation of the optimal batch size was performed using the Progressive Image Compression. Each model underwent training for 100 epochs using the training dataset. Subsequently, they were assessed in a testing environment using a set of 64 images sourced from the natural images dataset, with 8 images sampled from each of the 8 classes. It was observed that the optimal performance was achieved with a batch size of 8, yielding an SSIM of 97.57\%.

\subsection{Compression Ratio and Transfer Time}

\begin{figure}[h]
    \centering
    \includegraphics[width=0.45\textwidth]{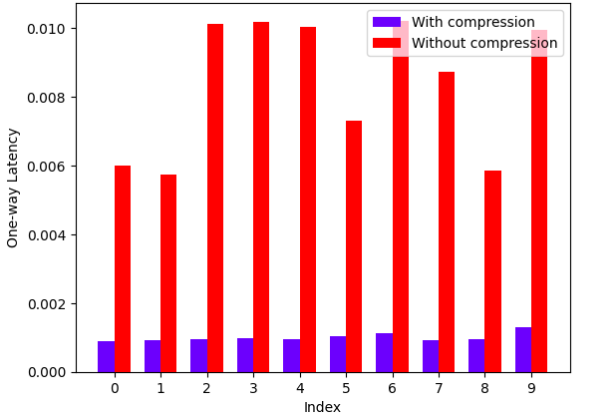}
    \caption{One-way transmission latency of compressed and uncompressed images}
    \label{fig:latency}
\end{figure}

Each input image was resized to a dimension of (256,256,3) to ensure compatibility with the encoder architecture. The compressed latent space had a dimension of (16,16,64). Thus the architecture was able to achieve a compression ratio of 12:1 concerning the resized images. 

To evaluate the effectiveness of the compression in a networked environment, we established a setup with two hosts connected via TCP sockets. Under identical network conditions, we transferred both the compressed and uncompressed images from one host to the other and recorded the one-way latency. The results of these transfers are illustrated in Figure \ref{fig:latency}. The evaluation revealed an average reduction in latency from uncompressed to compressed images of 87.5\%.

\subsection{Comparison with existing methods}
\begin{figure}[h]
    \centering
    \begin{subfigure}{0.42\textwidth}
        \centering
        \includegraphics[width=\linewidth]{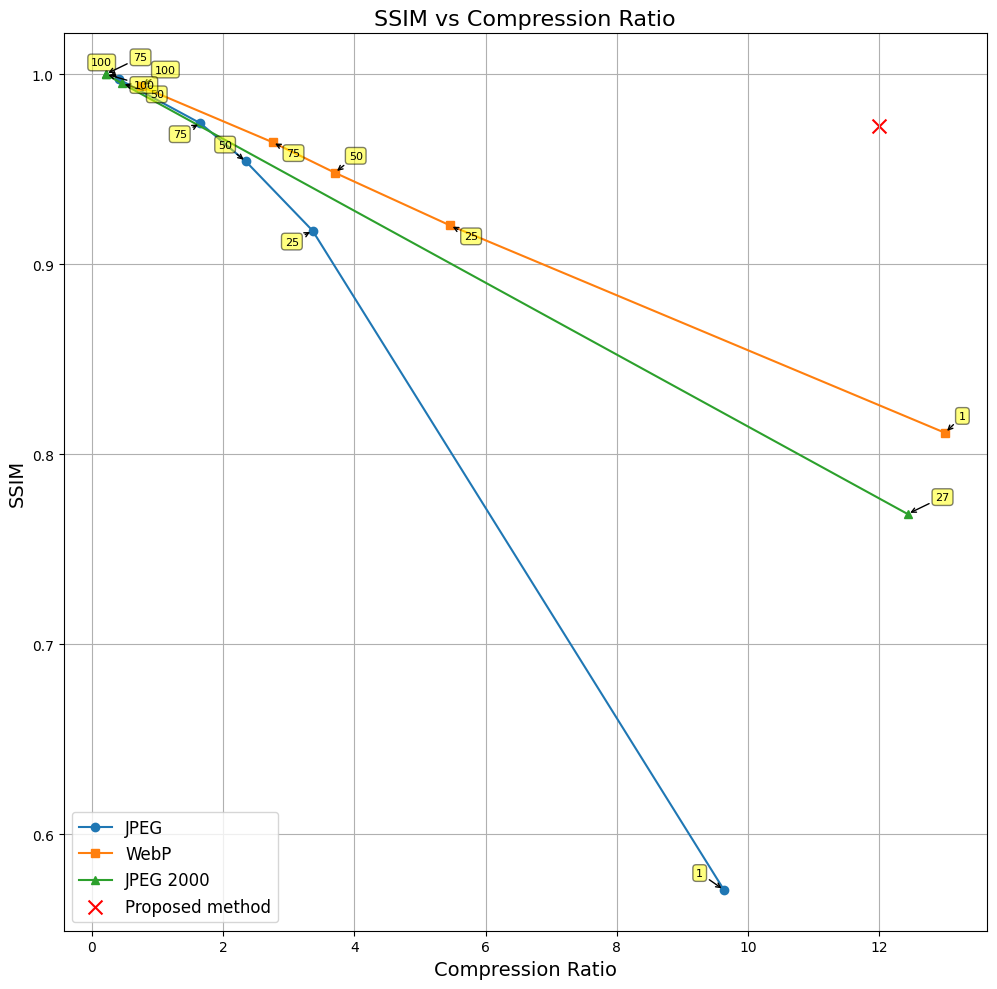}
        \caption{Structural Similarity Index Measure vs Compression Ratio}
        \label{fig:with_PgIC}
    \end{subfigure}%
    \hfill
    \begin{subfigure}{0.42\textwidth}
        \centering
        \includegraphics[width=\linewidth]{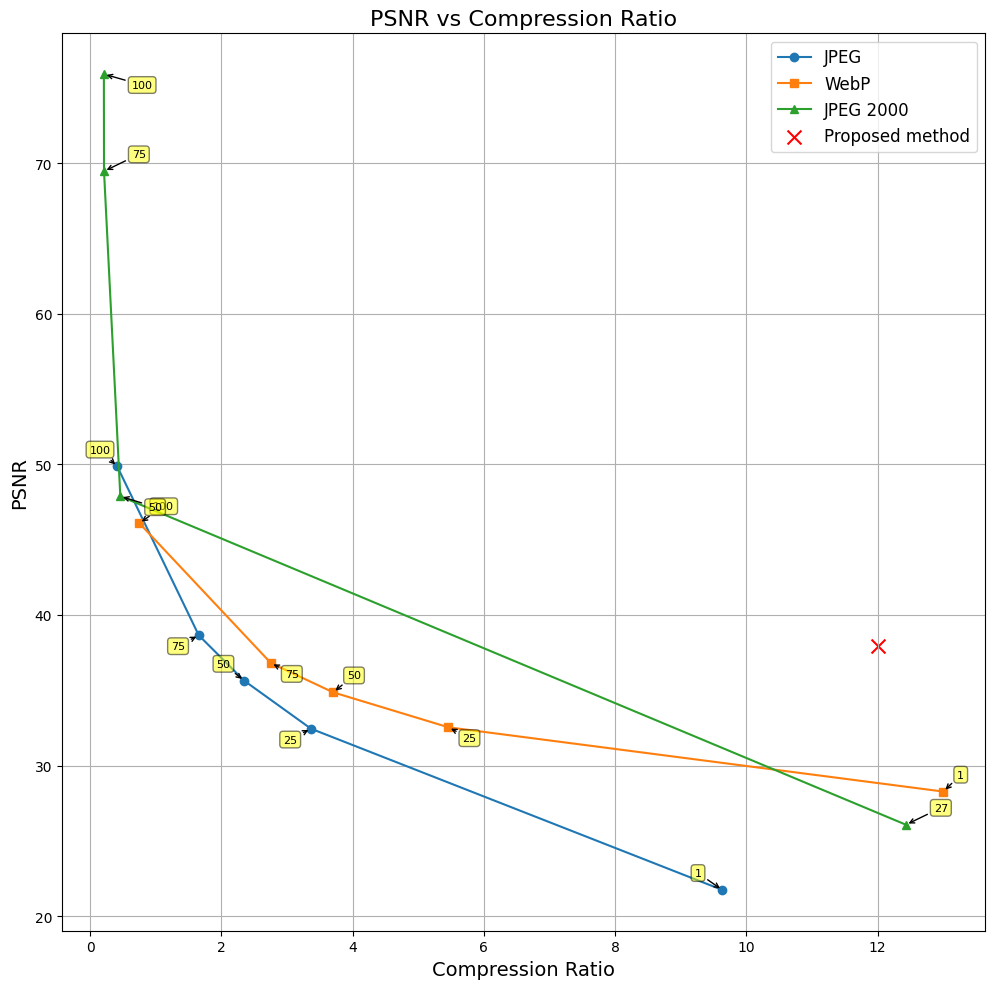}
        \caption{Peak Signal-to-Noise Ratio vs Compression Ratio}
        \label{fig:without_PgIC}
    \end{subfigure}
    \caption{Structural Similarity Index Measure and Peak Signal-to-Noise Ratio vs Compression Ratio values for various image compression techniques}
    \label{fig:comparison}
\end{figure}

Existing image compression methods typically achieve a compression ratio of about 2:1 while maintaining a high Structural Similarity Index (SSIM) of around 98\% to 99\%, ensuring minimal loss in image quality. In contrast, our proposed autoencoder not only achieves a comparable SSIM of 97.5\% but also significantly outperforms traditional methods with a compression ratio of 12:1. Additionally, the autoencoder incorporates encryption, providing enhanced security for the compressed images. We compared our autoencoder to the JPEG, WebP and JPEG 2000 algorithms across various quality settings. We analysed the trade-off between compression ratio and image quality metrics (PSNR and SSIM) for different quality settings. The results show that our proposed method achieves comparable PSNR and SSIM values to JPEG at a quality setting of 75 but with a six times higher compression ratio. The detailed comparison is illustrated in Figure \ref{fig:comparison}.
\section{Conclusion and Future Work}
The proposed solution of using convolutional autoencoders for image compression and its resultant encryption demonstrates significant results in maintaining image quality while ensuring secure transfer. The paper demonstrates how convolutional autoencoders effectively reduce image size by encoding to latent space, thereby preventing unauthorised access during transmission. The architecture was designed with considerations for PSNR, SSIM, and MSE, which are metrics that gauge the quality of reconstructed images. The idea of using a specific batch size for final training yielded optimal results for stability and reconstruction quality. Additionally, employing a progressive approach with residual images allowed the model to better reconstruct the image with lower training times.

The scope for future work lies in the areas of quantisation and entropy coding. Quantisation involves mapping a range of values into a single value, while entropy coding relies on the occurrence of data and the number of bits required to represent them. Using the PyAC library for entropy coding was a technique used in a few other papers and studies. Addressing these challenges and integrating these techniques could further enhance the efficiency and performance of the proposed autoencoder architecture.



\balance
\begingroup
\setstretch{0.75}
\bibliographystyle{ieeetr}
\bibliography{references}

\begin{thebibliography}{10}

\bibitem{b11}
E.~Johnson and F.~Brown, ``Enhancing autoencoder performance with advanced techniques,'' {\em arXiv [cs.CV]}, 2019.

\bibitem{b9}
F.~Osuolale, ``Secure data transfer over the internet using image cryptosteganography,'' {\em International Journal of Scientific and Engineering Research}, vol.~8, p.~1115, December 2017.

\bibitem{b15}
K.-L.~M. Takanori~Fujiwara, Oh-Hyun~Kwon, ``Supporting analysis of dimensionality reduction results with contrastive learning,'' {\em arXiv:1905.03911}, 2019.

\bibitem{b10}
C.~Smith and J.~Doe, ``Advanced techniques in image compression,'' in {\em Proceedings of the Australasian Conference on Artificial Life and Computational Intelligence}, 2018.

\bibitem{b1}
L.~Theis, W.~Shi, A.~Cunningham, and F.~Husz{\'a}r, ``Lossy image compression with compressive autoencoders,'' in {\em International conference on learning representations}, 2022.

\bibitem{b2}
Z.~Cheng, H.~Sun, M.~Takeuchi, and J.~Katto, ``Deep convolutional autoencoder-based lossy image compression,'' in {\em 2018 Picture Coding Symposium (PCS)}, pp.~253--257, IEEE, 2018.

\bibitem{b3}
S.~Petscharnig, M.~Lux, and S.~Chatzichristofis, ``Dimensionality reduction for image features using deep learning and autoencoders,'' in {\em Proceedings of the 15th international workshop on content-based multimedia indexing}, pp.~1--6, 2017.

\bibitem{b22}
Y.~Mei, L.~Li, Z.~Li, and F.~Li, ``Learning-based scalable image compression with latent-feature reuse and prediction,'' {\em IEEE Transactions on Multimedia}, vol.~24, pp.~4143--4157, 2021.

\bibitem{b4}
A.~Khosla, N.~Jayadevaprakash, B.~Yao, and F.-F. Li, ``Novel dataset for fine-grained image categorization: Stanford dogs,'' {\em IEEE Conference on Computer Vision and Pattern Recognition (CVPR)}, 2011.

\bibitem{b5}
C.~Alessio, ``Animals-10 dataset.'' \url{https://www.kaggle.com/datasets/alessiocorrado99/animals10}, 2020.

\bibitem{b6}
J.~Deng, W.~Dong, R.~Socher, L.-J. Li, K.~Li, and L.~Fei-Fei, ``Imagenet: A large-scale hierarchical image database,'' {\em IEEE Computer Vision and Pattern Recognition (CVPR)}, 2019.

\bibitem{b7}
S.~G. Prasun~Roy, SaumikBhattacharya, ``Natural images dataset.'' \url{https://www.kaggle.com/datasets/prasunroy/natural-images}, 2018.

\bibitem{b20}
B.~Lim, S.~Son, H.~Kim, S.~Nah, and K.~Mu~Lee, ``Enhanced deep residual networks for single image super-resolution,'' in {\em Proceedings of the IEEE conference on computer vision and pattern recognition workshops}, pp.~136--144, 2017.

\bibitem{b21}
K.~Li, S.~Yang, R.~Dong, X.~Wang, and J.~Huang, ``Survey of single image super-resolution reconstruction,'' {\em IET Image Processing Volume 14 Issue 11}, 2020.

\bibitem{b8}
Z.~Wang, A.~Bovik, H.~Sheikh, and E.~Simoncelli, ``Image quality assessment: from error visibility to structural similarity,'' {\em IEEE Transactions on Image Processing}, vol.~13, no.~4, pp.~600--612, 2004.

\bibitem{b12}
K.~Berahmand, F.~Daneshfar, E.~S. Salehi, Y.~Li, and Y.~Xu, ``A comprehensive survey on image classification techniques,'' {\em Artifcial Intelligence Review (2024)}, vol.~36, no.~6, pp.~615--638, 2024.

\bibitem{b13}
R.~G. Dor~Bank, Noam~Koenigstein, ``Autoencoders,'' {\em arXiv:2003.05991v2 [cs.LG]}, 2021.

\bibitem{b14}
A.~Brown and B.~Williams, ``Transforming auto-encoders,'' {\em Artificial Neural Networks and Machine Learning}, 2011.

\bibitem{b16}
H.~v. d.~H. L.J.P. van~der Maaten, E.O.~Postma, ``Dimensionality reduction: A comparative review,'' {\em Journal of Machine Learning Research 10(1)}, 2017.

\bibitem{b17}
S.~I. T.~j. S.Velliangiri, S.Alagumuthukrishnan, ``A review of dimensionality reduction techniques for efficient computation,'' {\em International Conference on Recent Trends in Advanced Computing}, 2019.

\bibitem{b18}
J.~L. . S.~H. Weikuan~Jia, Meili~Sun, ``Feature dimensionality reduction: a review,'' {\em Complex \& Intelligent Systems (2022) 8:2663–2693}, 2022.

\bibitem{b19}
A.~Bardos, I.~Mollas, N.~Bassiliades, and G.~Tsoumakas, ``Local explanation of dimensionality reduction,'' {\em arXiv:2204.14012v1 [cs.LG]}, 2022.

\end{thebibliography}
\endgroup

\end{document}